\documentclass{aa} 
  
\usepackage{color}
\usepackage{amsmath}
\usepackage{float}
\usepackage{natbib}
\usepackage{graphicx}

\usepackage{graphicx}
\usepackage{ulem}
\usepackage[varg]{txfonts}
\definecolor{magenta}{rgb}{1.0, 0.0, 1.0}
\definecolor{mygray}{gray}{0.6}

\newcommand{\eq}[1]{Eq.~(\ref{eq:#1})}

\def\me{$\,{\rm M}_{\oplus}\,$}
\def\re{$\,{\rm R}_{\oplus}\,$}
\def\h2o{H$_2$O}
\def\sio2{SiO$_2$}
\def\gc3{g\,cm$^{-3}$}

\usepackage{xspace}
\def\se{sub-Neptune\xspace}
\def\ses{sub-Neptunes\xspace}
\def\ssn{Sub-Neptune\xspace}

\begin{document}

\title{Effect of Core Cooling on the Radius of Sub-Neptune Planets}
\author{A.~Vazan\inst{1},
C.W.~Ormel\inst{1}, 
C. Dominik\inst{1}}
\institute{Astronomical Institute Anton Pannekoek, University of Amsterdam, The Netherlands.\\\label{inst1}
\email{a.vazan@uva.nl}
}

\abstract
{\ssn planets are very common in our galaxy and show a large diversity in their mass-radius relation. 
In \ses most of the planet mass is in the rocky part (hereafter {\it core}) which is surrounded by a modest hydrogen-helium envelope. 
As a result, the total initial heat content of such a planet is dominated by that of the core. 
Nonetheless, most studies contend that the core cooling will only have a minor effect on the radius evolution of the gaseous envelope, because the core's cooling is in sync with the envelope, \textit{i.e.}, most of the initial heat is released early on timescales of $\sim$10-100\,Myr. 
In this Letter we examine the importance of the core cooling rate for the  thermal evolution of the envelope. 
Thus, we relax the early core cooling assumption and present a model where the core is characterized by two parameters: the initial temperature and the cooling time.
We find that core cooling can significantly enhance the radius of the planet when it operates on a timescale similar to the observed age, \textit{i.e.} $\sim$Gyr.
Consequently, the interpretation of \ses' mass-radius observations depends on the assumed core thermal properties and the uncertainty therein.
The degeneracy of composition and core thermal properties can be reduced by obtaining better estimates of the planet ages (in addition to their radii and masses) as envisioned by future observations.}

\keywords{Methods: numerical, Planetary systems, Planets and satellites: composition, Planets and satellites: interiors, Planets and satellites: physical evolution}

\maketitle

\section{Introduction}\label{intro}
Super-Earths and mini-Neptunes (hereafter {\it \ses}) -- planets with masses between terrestrial planets and ice giant planets -- are a very common type of planets of our galaxy \citep[\textit{e.g.},][]{howard12,batlaha13,coughlin16}.
Many of these planets {are inferred to} contain some amount of hydrogen and helium, typically several percent by mass \citep{lopez12,jontof16,wolfglopez15,fulton17}, and show a large scatter in their observed mass-radius (MR) {relation}.
Usually {the MR-scatter} is interpreted in terms of a compositional diversity, where planets of a larger radius (for the same mass) are assigned a larger fraction of hydrogen and helium \citep[\textit{e.g.},][]{lopezfor14}.

However, the planet radius is time-dependent and is affected by the thermal evolution of the planet. 
The radius evolution of the rocky part of the planet (hereafter {\it core}\footnote{Our definition of \textit{core} includes the mantle.}) is only weakly dependent on its composition, due to degeneracy in heavy elements properties \citep{rogersseag10}.
A much larger indirect effect of the core on the radius of the planet occurs when the core is covered by a hydrogen-helium (HHe) envelope of only a few percent in mass, as is the case for many \se planets. 
Hence, for \se planets, most of the mass is in the core, while the radius of the planet is determined by the entropy of the HHe-dominated envelope.

The energy content of the core is usually expected to be smaller than the envelope's for planets with a large fraction of HHe due to the lower heat capacity of rocks in comparison to hydrogen and helium, as derived from equations of state. 
But for planets with an envelope mass of a few percent most of the energy is in the core \citep{ginzburg16}. 
Moreover, the initial core energy content -- a property determined by the planet formation process -- can be much higher than that of the envelope, and raise the importance of the core as an energy reservoir for the envelope.
As a result, of any planet type we can expect the radius evolution of gas-rich \se planets to be most affected by the underlying thermal evolution of its rocky core.

In most works \citep[\textit{e.g.}][]{lopez12,nettel11,chenrog16} the thermal evolution of the core is not followed explicitly; instead, the core cooling rate, ${dT_\mathrm{core}}/{dt}$ is simply taken to be identical to the  cooling at the base of the envelope ($T_\mathrm{env}$), such that
the luminosity generated by core cooling is
\begin{equation} \label{Lc_simple}
  L_\mathrm{core} = c_v M_c \frac{dT_\mathrm{core}}{dt} = c_v M_c \frac{dT_\mathrm{env}}{dt}
\end{equation}
where $c_v$ is the core specific heat capacity, $M_c$ the core mass. 
Implicitly, the assumption entering Eq.\ (\ref{Lc_simple}) is that core cooling is as fast as the cooling of the adiabatic envelope, \textit{i.e.}, determined by the radiative cooling rate at the outer atmosphere.
Under this assumption (hereafter {\it standard} approximation) the long-term thermal evolution is not affected by core properties \citep{chenrog16}.
But the core cooling timescale may be different than that of the envelope. If the envelope cools faster than the core, the core contribution becomes dominant after the envelope has cooled. Thus, the role of the core in the thermal evolution depends on its cooling timescale.

Since most of the detected planets orbit main-sequence field stars (\textit{i.e.,} ages of 1-10\,Gyr), the core cooling timescale is a key parameter for the interpretation of observed radii by structure evolution models of \ses.
In this Letter we examine the importance of the core cooling rate for the thermal evolution of the envelope; that is we relax the assumption that the core heat transport is as rapid as that of the envelope.
We introduce a parameter $t_\mathrm{cool}$ for the timescale of core cooling and investigate the consequences of short and long core cooling timescales on the radius evolution of the planet.

\section{Model}
When the core cooling operates on a timescale longer than that of the envelope, the standard assumption no longer applies. 
To account for these effects, we consider an alternative expression for the core luminosity:
\begin{equation}\label{eq:Lc_new}
  L_\mathrm{core} = c_v M_c \,max\left(\frac{dT_\mathrm{env}}{dt}\,, \frac{dT_\mathrm{*core*}}{dt}\right)
\end{equation}

In Eq.\ (\ref{eq:Lc_new}) the core luminosity is taken to be the maximum between the natural core cooling rate, $dT_\mathrm{*core*}/dt$, and the envelope cooling rate, $dT_\mathrm{env}/dt$, set by the radiative cooling through the atmosphere.\footnote{We do keep the $dT_\mathrm{env}/dt$ term to prevent decrease of the core temperature below the (base of the) envelope's temperature.}

In this exploratory work, the natural core contribution by $dT_\mathrm{*core*}/{dt}$ is highly simplified, being determined by two parameters only: the core initial temperature, $T_{c,0}$, which reflects the initial core energy content, and the timescale for core cooling, $t_\mathrm{cool}$, which regulates how fast heat is transported between core and the envelope:
\begin{equation}  \label{eq:Tc_new}
T_\mathrm{*core*} = T_{c,0}\exp\left( -\frac{t}{t_\mathrm{cool}}\right)
\end{equation}

All uncertainties regarding initial core heat contents and core heat transport are encapsulated in {these two} parameters.
We insert \eq{Tc_new} in \eq{Lc_new} and solve for the core luminosity as function of time:
\begin{equation} \label{eq:Lc_our}
L_\mathrm{core} = c_v M_c \, max\left(\frac{dT_\mathrm{env}}{dt}\,, \frac{T_{c,0}}{t_\mathrm{cool}} \exp\left(-\frac{t}{t_\mathrm{cool}}\right)\right)
\end{equation}

The envelope cooling timescale of ${dT_\mathrm{env}}/{dt}$ term in \eq{Lc_our} is on the order of several $10^7$\,yr for solar metallicity opacity \citep[e.g.,][]{lopezfor14,chenrog16}. 
Thus, the contribution of the initial energy content of $dT_\mathrm{env}/{dt}$ is negligible in comparison to $dT_\mathrm{*core*}/{dt}$ for timescales longer than $10^7$\,yr.
Concerning the timescale $t_\mathrm{cool}$, we identify three regimes:
\begin{itemize}
    \item $t_\mathrm{cool}\ll t$: the timescale for heat release from the core is much shorter than the current age; the core has already cooled to the atmosphere's temperature by the time $t$, and thus the contribution of the core luminosity ${dT_\mathrm{*core*}}/{dt}\rightarrow 0$. As a result, the evolution is determined entirely by the cooling of the envelope (${dT_\mathrm{env}}/{dt}$) which is the standard approximation. 
    \item  $t_\mathrm{cool}\sim t$: the age of the system is about equal to the timescale of core cooling; in this case the significant heat flux from the core cooling keeps the envelope extended. Hence, we get the maximal effect of the core luminosity on the envelope thermal evolution.
    \item $t_\mathrm{cool}\gg t$: the core cooling timescale is much longer than the planet age. Most of the heat is still locked up in the core, but the core luminosity is reduced by ${T_{c,0}}/{t_\mathrm{cool}}\rightarrow 0$. Therefore, the ${dT_\mathrm{*core*}}/{dt}\rightarrow 0$ and the evolution is determined again by the envelope.
\end{itemize}

We apply our models for planets in the sub-Neptune mass regime.  
The envelope mass fraction increases with mass \citep{mordasini12}; here we take 3, 5, 10\me planets with envelope masses of 1\%, 10\%, 15\%, respectively. 
We consider different core cooling timescales $t_\mathrm{cool}$, while the envelope cooling time is not constrained.

The initial core temperature $T_{c,0}$ is {estimated from the gravitational binding} energy of the core.
For a core of mass $M_c$ and radius $R_c$:
\begin{equation}\label{eq:tmax}
    T_\mathrm{max} \sim\frac{G{M_c}}{R_c C_p}
\end{equation}
This assumes no radiative or advective losses. 
The fraction of the gravitational binding energy which remains in the core in the form of thermal energy is determined by the solid accretion rate during core formation.
Since we do not specify a model for the core formation, we consider an arbitrary value of $\sim20\%$ of the binding energy to remain in the core after formation.
We calculate the core radii by using our rock (\sio2) equation of state \citep{vazan13}. The resulting initial core temperatures for the planets in our model are $2.5\times10^4$\,K, $3.1\times10^4$\,K, and $5.6\times10^4$\,K respectively.

The evolution is calculated by a 1D hydrostatic planetary evolution code \citep{vazan13} that solves the structure and evolution equations for the rocky (\sio2) core and hydrogen-helium envelope.
We consider the core luminosity as in \eq{Lc_our}, embedded in a gaseous envelope of hydrogen and helium in a solar ratio. 
The initial envelopes are adiabatic, with entropy of $s=9\,\mathrm{k_B/baryon}$. 
In order to isolate the effect of core cooling on the radius evolution we minimized other thermal effects of opacity, irradiation and photo-evaporation: 
(1) radiative opacity determines the atmosphere heat transport and therefore the cooling of the planet. High metallicity opacity will slow the planet cooling and keep the planet radius larger for longer. Here
we use radiative opacities of \cite{sharp07} for solar metallicity (non-enhanced) planetary atmospheres.
(2) irradiation by the parent star is not included in the model. 
Stellar irradiation is expected to slow the planet cooling, and therefore enhance core thermal evolution effects \citep{baraffe08}.
(3) The planet mass is constant during the evolution, \textit{i.e.}, no photo-evaporation \citep{owenwu13} or accretion are included in the model.

\section{Results}
\begin{figure}[t]
    \centerline{\includegraphics[width=8.5cm]{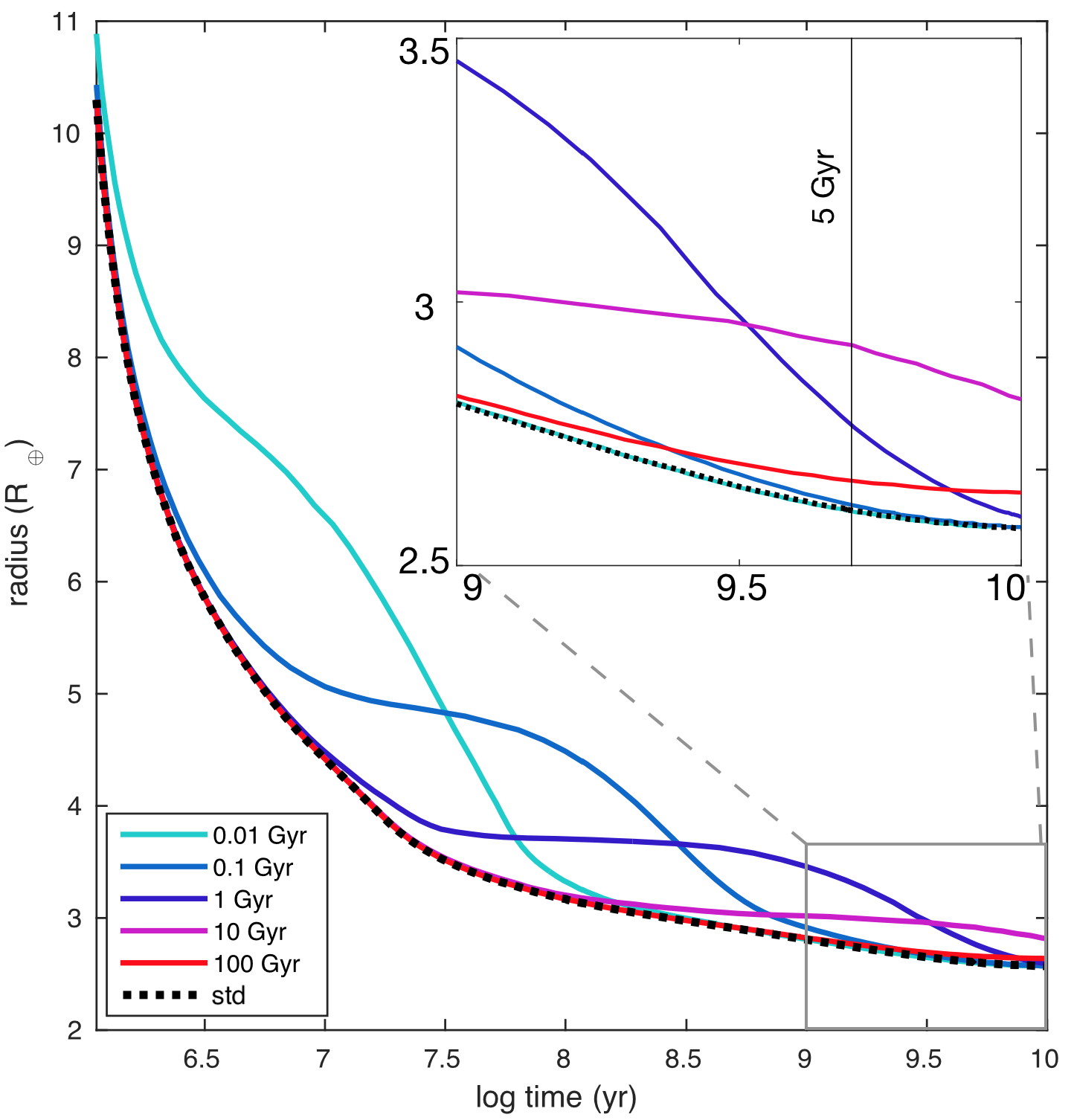}}
    \caption{Radius evolution of a 5\me planet with 10\% HHe envelope for different core cooling times. The dashed curve is for standard core cooling rate approximation. The initial conditions are the same for all cases. In the rectangle: radius evolution between 1-10\,Gyr.}\label{fg:Rev}
\end{figure}

To validate our model we have calculated the thermal evolution of \se planets under the standard approximation, (${dT_\mathrm{*core*}}/{dt}=0$) and have compared our results to \cite{baraffe08} and \cite{lopezfor14}. We find convergence in the planet's radii between the models with differences at the 2\% level.
\footnote{To conduct a proper comparison, we adjusted our model in order to match the setup of the \cite{baraffe08} (where the core is made of \h2o) and \cite{lopezfor14} (where stellar irradiation is included).}

Then, we add the core luminosity ${dT_\mathrm{*core*}}/{dt}$ as described in \eq{Lc_our}, for five (order of magnitude) core cooling timescales $t_\mathrm{cool}$ between $1\times10^8$ and $1\times10^{11}$\,yr. 
In Fig.~\ref{fg:Rev} we present the radius evolution of a 5\me planet with a rocky core and a 10\% HHe envelope. The different colors correspond to the different core cooling timescales and the dotted curve corresponds to the standard approximation of core cooling on envelope cooling ratio (${dT_\mathrm{*core*}}/{dt}=0$). As is shown in Fig.~{\ref{fg:Rev}}, the radius evolution is clearly affected by the core cooling timescale. 
When we focus on the radius evolution in the typical observation timescale (grey rectangle zoom in), we find significantly different radii for the different models.
In general, the radius evolution is similar to the standard case whenever the energy flux from the core is minor. On the core cooling timescale the envelope remains extended as long as the core energy is being released. When the heat flux from the core diminishes the planet contracts to the radius of the standard case.
Thus, the greater change in radius occurs right after the core cooling time. 

Therefore, if the core cools on a timescale similar to the observed age (\textit{e.g.}, purple and magenta curves in Fig.~\ref{fg:Rev}) the core cooling effect on the radius is maximal.
If the core cools on 1\,Gyr time (purple curve), the change in radius from 1\,Gyr to 10\,Gyr is about 30\%. This is much larger than the change in the adiabatic evolution case (dotted curve) within the same time period. 
For a core cooling timescale of 10\,Gyr (magenta curve) the planet radius is $\sim 15\%$ larger than the radius of the standard case, through most of the evolution between 1-10\,Gyr. 
On the other hand, planets with core cooling timescales much shorter than 1-10\,Gyr (cyan and light blue curves) have already released most of their core {heat contents by $\sim$1 Gyr, rendering} its radius evolution similar to the standard case.
Finally, when the core cooling time is much longer than the typical observation age (red curve) the core stays hot but its thermal effect on the radius is small by $\sim$10 Gyr. 

This calculation has been conducted for a particular choice (20\%) of the initial energy content as derived from \eq{tmax}. 
Higher (lower) initial core temperatures increase (decrease) the core thermal effect on the radius.

\begin{figure}[t]
    \includegraphics[width=8cm]{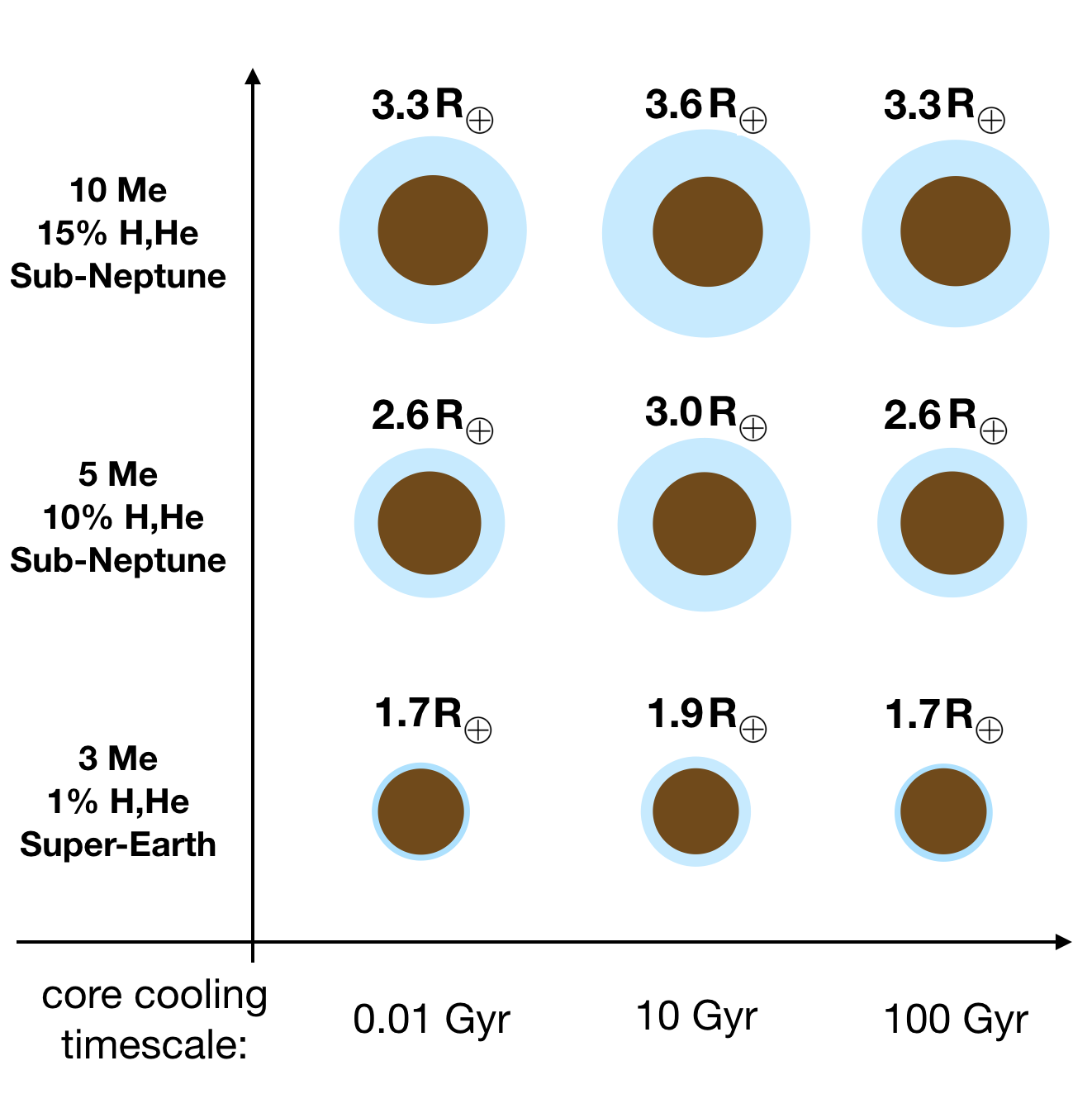}
    \caption{Core and envelope radii at age of 5\,Gyr for diferent planet types and different values of $t_\mathrm{cool}$. The radii are to scale. Once the core cooling time is similar to the planet age (\textit{i.e.} middle column) the liberated heat causes the radius to expand.}\label{crtn}
    \end{figure}

In Fig.~\ref{crtn} we illustrate planet and core radii after 5\,Gyr for different planetary masses and values of core cooling times.
We find the differences in radius between core cooling timescale of 10\,Gyr and the standard models to be 11\%, 15\%, 9\% respectively for this age (5\,Gyr). 
As the planet mass increases or decreases from the \se mass range, the effect of the core cooling becomes weaker: 
more massive planets feature a stronger gravity, which diminishes the importance of thermal effects on the envelope. In addition, higher HHe mass fractions in more massive planets lower the core-to-envelope mass ratio and therefore reduce the core contribution.
On the other hand, the small amounts of HHe that lower mass planets contain imply that their total radius is dominated by that of the core, rendering core and envelope thermal effects insignificant.
Hence, we find the maximal effect for the 5\me planet in our model.

In Table~\ref{tab2} we provide radii for the different masses in this work for planet ages of 1, 5, and 10\,Gyr.   
In the bottom part of the table we present radii for 5\me and 10\me planets with different envelope mass fractions than above.
As was shown for 5\me in Fig.~\ref{fg:Rev}, a short cooling time of 0.01\,Gyr and a long cooling time of 100\,Gyr result in radii very similar to the standard core cooling between 1-10\,Gyr. Therefore we do not include them in the table, but add the standard case instead.
The radius trends we find for 5\me planets appears also for the 3\me and 10\me planets, but are more moderate.

The maximum we find for core thermal effects in 5\me planet is valid when the mass of the envelope is increasing with the mass of the core. For independent envelope mass (e.g., same envelope fraction for different core mass) the effect of the core on the radius evolution increases as the planet mass decreases, due to lower gravity.
As shown in the bottom part of Table~\ref{tab2}, a 5\me planet with 1\% HHe envelope exhibits smaller change in radius in comparison to the 3\me planet with the same 1\% HHe (upper table), for the same cooling properties. Similar trend operates for a 10\% HHe envelope of a 5\me planet (upper table) in comparison to a 10\me planet (lower table).

\begin{table}[t]
\centering 
\begin{tabular}{c c c c c} 
 \hline\hline\\[-1.5ex] 
mass [\me] & $t_\mathrm{core}$[Gyr] & & radius [\re] & \\         
\hline\\[-1.5ex]
 & & at 1\,Gyr & at 5\,Gyr & at 10\,Gyr \\         
\hline\\[-1.5ex]
3 & std & 1.76 & 1.71 & 1.7\\
3 & 0.1 & 1.82 & 1.71 & 1.7\\
3 & 1 & 2.13 & 1.76 & 1.71\\
3 & 10 & 1.91 & 1.87 & 1.82\\
5 & std & 2.81 & 2.6 & 2.57\\
5 & 0.1 & 2.91 & 2.6 & 2.57\\
5 & 1 & 3.45 & 2.76 & 2.6\\
5 & 10 & 3.1 & 2.98 & 2.86\\
10 & std & 3.47 & 3.28 & 3.23\\
10 & 0.1 & 3.55 & 3.29 & 3.23\\
10 & 1 & 4.04 & 3.47 & 3.28\\
10 & 10 & 3.7 & 3.64 & 3.53\\
\hline 
\\[-1ex]
5 (1\%)& std & 1.96 & 1.88 & 1.85\\
5 (1\%)& 1 & 2.28 & 1.95 & 1.85\\
10 (10\%)& std & 3.2 & 3.04 & 3.03\\
10 (10\%)& 1 & 3.69 & 3.21 & 3.04\\
 \hline 
  \\[0.01ex] 
\end{tabular}
\caption{Planet radii for different planetary masses and cooling timescales. The std $t_\mathrm{cool}$ is for the standard approximation of core cooling on the envelope's rate. The initial energy content for all cases presented is taken as 20\% of the binding energy. No irradiation by the star is included. The bottom part of the table is for different envelope fraction (in parenthesis) than in the models above.
}\label{tab2}
\end{table}

\section{Discussion}
In this Letter we have demonstrated that {protracted} core cooling can, under certain conditions, significantly {increase} the {radii} of \se planets {on timescales of $\sim$Gyr}. 
Therefore, the conversion of observed radius-mass relation into envelope mass fraction actually gives an upper bound to the HHe mass fraction when the standard core cooling approximation is adopted, as is common. 
The diversity in mean density of \ses is not determined solely by composition (\textit{i.e.}, hydrogen and helium fraction), but can also be driven by a different formation history and by a diversity in the cooling timescale(s) of planet cores.

What, then, are realistic value for $T_{c,0}$ and $t_\mathrm{cool}$? 
Broadly, energy transport in the core will be either convective or conductive, with the former being able to support much larger heat fluxes than the latter. 
Convective heat transport is {determined by} the Rayleigh number, which depends on structure properties (gravity, density, layer length and temperature profile), but also on material properties such as thermal expansion coefficient, thermal diffusivity and kinematic viscosity. 
For modellers, the problem is the large uncertainty of these properties as a function of temperature and pressure \citep[e.g.,][]{valencia06,tachinami11,vandenBerg10,stamenkovic12}. 
Within the uncertainly limits, \citet{stamenkovic12} for example, show that different dependencies of viscosity on pressure provide various core cooling timescales. 
In \citet{vandenBerg10} the authors suggest that conductivity in high pressure conditions can make conduction favourable over convection in the long term. 
Thus, mechanisms that operated for Earth-like conditions may not be equally applicable to \se pressure-temperature conditions. 
Conductive heat transport, on the other hand, can be estimated by $t_{\rm cond}\sim{\rho\, C_p\,  R_c^2}/{\kappa_\mathrm{cond}}$.
For conductive opacities $k_\mathrm{cond}$ from \cite{potekhin99}, the conductive timescale is of the order of $10^{10}-10^{12} \rm yr$.
Given their implications on the radius evolution of \ses, as outlined by this Letter, a closer understanding of the core cooling properties is clearly an important subject for further investigation.

In our calculations in Section~3 we considered that the thermal energy content of the core was solely due to its gravitational binding energy, where we adopted an efficiency of 20\%. 
This amounts to a specific energy of $\sim10^{11}$\,erg/g.
However,
additional heat generation mechanisms can operate in the core, and increase the core energy content. 
Release of latent heat in phase transitions depends on the evolution of the core pressure-temperature profile. 
Latent heat for \sio2, for example, in liquid-solid transition is $\sim6\times10^9$\,erg/g \citep{richet82}.
Radiogenic heating also contribute the core energy during the evolution. 
The dominant radioactive elements $\rm U^{\rm 238},U^{\rm 235},Th^{\rm 232}$ and $K^{\rm 40}$ have half lives in the Gyrs regime and provide $\sim3\times10^{10}$\,erg/g for meteorite-like composition \citep{grevesse93,nettel11} .
{Similarly,} core contraction \citep{baraffe08} is an additional contributor to the energy balance of the core.
Given these additional heat-generating mechanisms, core thermal effects on the radius evolution can be enhanced. 

The model presented in this Letter is a first step {towards a more complete} model of the simultaneous evolution of core and envelope.
In order to identify the core thermal contribution and to constrain the structure and composition of these planets, the core should be modelled in more detail. 
{Indeed,} a proper model for $L_\mathrm{core}$ entails that we {include} the core {in a} center-to-surface model together with the envelope. 
In a future study (Vazan et al. in prep) we will investigate the effect of various thermal parameters on the radius evolution by {such} multi-zone model for the core.

Our results reveal a significant degeneracy in the mass-radius relation between composition (mainly core to envelope mass ratio) and core cooling behaviour. 
This degeneracy takes place when the core cooling timescale is similar to the age at which we observe the planet.
Future observations can, however, help to break this degeneracy by {a} better estimation of the planet (or stellar) ages, as well as more {precise} mass-radius {observations}.
Our results indicate that the planet radius evolution can exhibit a characteristic trend {over $\sim$Gyr} timescales, which is a direct consequence of the core cooling behaviour ($t_\mathrm{cool}$). 
Hence, if the temporal dimension is added, a statistical analysis of {observed} mass-radius-age {data} for \se planets can constrain core cooling models and the core properties. 
For example, several planets in the sub-Neptune mass range with extended radii ("super-puffs"), like Kepler-51b,c \citep{steffen13,masuda14} or Kepler-79d \citep{jontof14} are planets in systems younger than 3.5\,Gyr. 
Therefore, if {the} typical core cooling timescale is {of} the order of 1\,Gyr, {a} much more modest HHe fraction would be required to explain their radii, than what is currently suggested.

\begin{acknowledgements}
We thank the anonymous referee for constructive comments that improved this manuscript. A.V.\ acknowledges support by the Amsterdam Academic Alliance (AAA) Fellowship.
C.W.O.\ is supported by the Netherlands Organization for Scientific Research (NWO; VIDI project 639.042.422)
\end{acknowledgements}

\bibliographystyle{aa}
\bibliography{allona}

\end{document}